%% file: main.tex
\documentclass[conference, a4paper,9pt]{IEEEtran}
\IEEEoverridecommandlockouts

\usepackage{amssymb}
\usepackage{cite, }
\usepackage{amsmath,amsfonts}
\usepackage{newtxmath}
\usepackage{graphicx}
\usepackage[dvipsnames]{xcolor}
\usepackage{textcomp}
\usepackage{threeparttable}
\usepackage{multirow}
\usepackage[linesnumbered,ruled,vlined]{algorithm2e}
\usepackage{svg}
\usepackage{caption}
\usepackage{subfigure}
\usepackage{hhline}
\usepackage{enumitem}
\usepackage{listings}
\usepackage{booktabs} 
\usepackage{tablefootnote}
\usepackage{afterpage}
\usepackage{colortbl}
\usepackage{soul}
\usepackage{amsmath, amssymb} % for \checkmark
\usepackage{pifont} % for \ding
\usepackage{afterpage}
\usepackage{float}
\usepackage{microtype}

\usepackage{graphicx}

\usepackage{booktabs}
\usepackage{multirow}
\usepackage{threeparttable}
\usepackage{array}
\usepackage{hyperref}
\usepackage{cleveref}
\crefname{section}{§}{§§}
\Crefname{section}{§}{§§}

\newcommand{\gemmchipname}{\mbox{\textit{Voltra}}\xspace}

\title{A 16 nm 1.60TOPS/W High Utilization DNN Accelerator with 3D Spatial Data Reuse and Efficient Shared Memory Access}
\author{Xiaoling Yi, Ryan Antonio, Yunhao Deng, Fanchen Kong, Joren Dumoulin, Jun Yin, Marian Verhelst\\
\normalsize{MICAS-ESAT, KU Leuven, Leuven, Belgium} \\
{\normalsize\{xiaoling.yi, ryan.antonio, yunhao.deng, fanchen.kong, joren.dumoulin, jun.yin, marian.verhelst\}@kuleuven.be}
}

\begin{document}

\maketitle

\vspace{-20em}
\input{content/0-abs}
\input{content/1-intro}
\input{content/2-chip_arch}
\input{content/3-exp}

\input{content/4-conclusion}
\input{content/5-ack}

\clearpage

% must specify the style
\bibliographystyle{IEEEtran}
\bibliography{ref}

\end{document}

%% file: content/0-abs.tex
\begin{abstract}
% With the rapid deployment of deep neural networks (DNNs) across diverse applications, there is an increasing demand for DNN accelerators that deliver both high energy efficiency and flexibility.
% To achieve high compute utilization across a wide range of convolutional neural network (CNN) and Transformer workloads, we propose a novel chip, named \gemmchipname, that leverages 3-Dimensional (3D) spatial data reuse along with efficient and flexible shared memory access. 
Achieving high compute utilization across a wide range of AI workloads is crucial for the efficiency of versatile DNN accelerators. This paper presents the \gemmchipname chip and its utilization-optimised DNN accelerator architecture, which leverages 3-Dimensional (3D) spatial data reuse along with efficient and flexible shared memory access.
The 3D spatial dataflow enables balanced spatial data reuse across three dimensions, improving spatial utilization by up to $2.0\times$ compared to a conventional 2D design.
Inside the shared memory access architecture, \gemmchipname incorporates flexible data streamers that enable mixed-grained hardware data pre-fetching and dynamic memory allocation, further improving the temporal utilization by $2.12$-$2.94\times$ and achieving $1.15$-$2.36\times$ total latency speedup compared with the non-prefetching and separated memory architecture, respectively.
% \gemmchipname further proposes a time-multiplexing technique to reduce module area cost while incurring minimal performance loss. 
Fabricated in 16nm technology, our chip achieves $1.60$ TOPS/W peak system energy efficiency and $1.25$ TOPS/mm\textsuperscript{2} system area efficiency, which is competitive with state-of-the-art solutions while achieving high utilization across diverse workloads.
\end{abstract}


%% file: content/1-intro.tex
\section{Introduction}
\label{sec:intro}

Artificial Intelligence (AI) has become an integral component of modern technology, fostering a proliferation of innovative applications that permeate daily life. However, this advancement has also introduced an insatiable demand for high-performance and energy-efficient deployment, particularly at the edge, where stringent power and area constraints prevail. At the same time, the rapid evolution and diversity of AI algorithms underscore the pressing need for flexibility in accelerator chip design. Thus, there is an increasing demand for DNN accelerators that deliver both high energy efficiency and flexibility.
% first the need for efficiency, then the need for flexibility

\begin{figure}[t]
	\centering
	\includegraphics[width=0.96\linewidth]{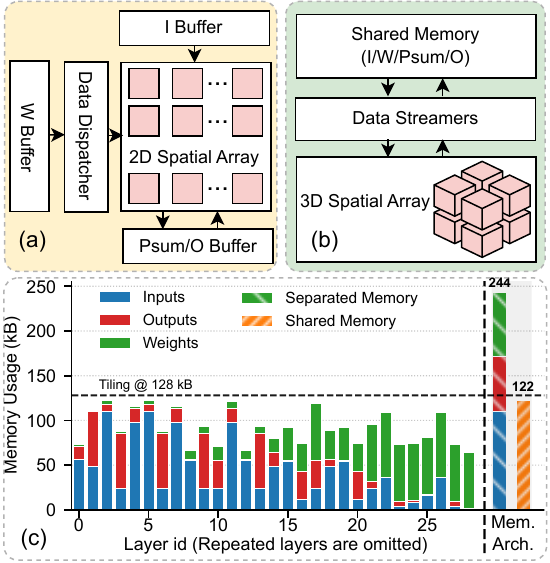}
	\caption{(a) 2D spatial array with separated data buffers and dispatchers. (b) 3D spatial array with shared memory and data streamers. (c) Memory usage comparison for the same tiling strategy of ResNet50 between the separated and shared memory.}
	\label{fig:motivation}
    \vspace{-0.6cm}
\end{figure}

% discuss diana, opengemm, huaa, RBE, RedMule, 
In the past decade, numerous contemporary domain-specific DNN accelerators~\cite{chen2016eyeriss, houshmand2022diana, du202328nm, tong2024feather, shi202516nm, zhou202523, feng2025aspen, sijstermans2018nvidia} have emerged, featuring extensive optimizations in spatial array design and dedicated scratchpad memories, achieving orders-of-magnitude performance and efficiency gains over traditional CPUs and GPUs. Despite employing intensive optimization strategies, these accelerators generally share a similar architectural template: a 2-Dimensional (2D) spatial array that enhances spatial data reuse, paired with dedicated, separate data buffers and dispatchers for each operand of the array to provide high data bandwidth, as illustrated in Figure~\ref{fig:motivation}~(a).

% the sota, first datapath, % then data access
However, as DNN workloads become increasingly diverse, the rigid 2D spatial array can suffer from low spatial utilization due to the mismatches between workload sizes and array dimensions \cite{shi2024bitwave, du202328nm, shi202516nm, yi2025opengemm, ghodrati2020planaria, kim2024dacapo}. OpenGeMM \cite{yi2025opengemm} addresses this issue by introducing a 3D spatial array that balances the unrolling of three computational dimensions and supports spatial accumulation, thereby improving spatial utilization across diverse workloads. Furthermore, the separated buffer architecture imposes significant constraints on workload tiling, where the on-chip dedicated buffers can not store all the operands for a single layer: \textit{the tiling strategy must conform to the size of the smallest buffer}. This prevents full utilization of the on-chip memory, and thus reduces data reuse possibilities and increases pressure on the off-chip memory communication. Shared memory architecture with data streamers \cite{yi2025datamaestro, genc2021gemmini, zaruba2020snitch, antonio2025open, tortorella2023redmule, cuyckens2025precision, kim2025maveric} address this limitation by enabling dynamic memory allocation across different operands. Figure~\ref{fig:motivation}~(b) presents an overview of a 3D spatial array with shared memory architecture, and Figure~\ref{fig:motivation}~(c) shows that the shared memory structure uses 50\% less memory for the same tiling strategy of the ResNet50 workload compared with a separated memory architecture. However, shared memory can introduce severe bank contention among simultaneous operand accesses, reducing the temporal utilization of the spatial array~\cite{yi2025datamaestro, colagrande2025towards}. 
% A mismatched data layout management and dedicated data manipulation unit would exacerbate this issue~\cite{yi2025datamaestro}.

% our contribution
In this work, we present \gemmchipname, a DNN accelerator featuring a General Matrix Multiplication (GeMM) core with a 3D spatial array and a shared data memory enhanced by flexible and efficient data streamers. Inside \gemmchipname, mixed-grained hardware data pre-fetching is employed to boost temporal utilization, while programmable dynamic data allocation enabled by the shared memory architecture and programmable data streamers effectively reduces off-chip data communication.
A time-multiplexing scheme significantly reduces module area with minimal performance loss. 
Comprehensive performance evaluation across diverse CNN, RNN, and transformer workloads shows that: (1) the 3D spatial array improves spatial utilization by up to $2.0\times$ over a 2D design; (2) mixed-grained data pre-fetching further enhances temporal utilization by $2.12$-$2.94\times$; and (3) the programmable dynamic data allocation improves overall latency by $1.15$-$2.36\times$ compared to separated memory structures when considering practical tiling and the off-chip memory access effects. Fabricated in 16nm technology, \gemmchipname achieves $1.60$ TOPS/W peak system energy efficiency and $1.25$ TOPS/mm\textsuperscript{2} system area efficiency, which is competitive with state-of-the-art designs while maintaining high utilization across varied workloads.

%% file: content/2-chip_arch.tex
\section{Chip Architecture}
\label{sec:arch}

% \subsection{Overview}
% \label{sec:arch_overview}

\begin{figure}[t]
	\centering
	\includegraphics[width=1\linewidth]{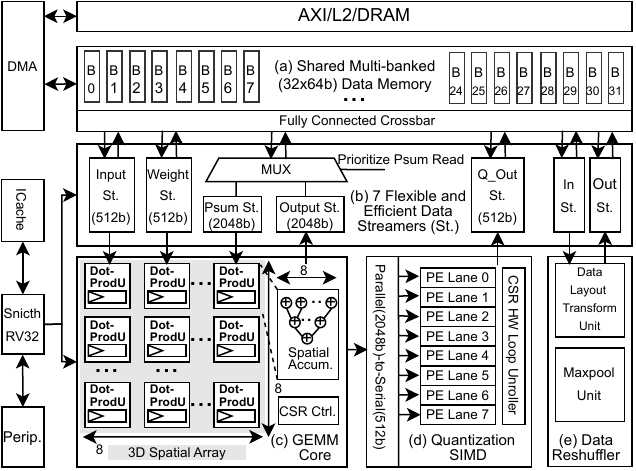}
	\caption{\gemmchipname architecture overview.}
	\label{fig:chip_arch}
    \vspace{-0.5cm}
\end{figure}

Figure \ref{fig:chip_arch} illustrates the overall architecture of \gemmchipname, which comprises three main functional blocks: a GEMM core, a quantization SIMD unit, and a data reshuffler. These blocks share a unified multi-bank memory structure (32 banks, 64-bit width each) that facilitates dynamic memory space partition. Flexible data streamers are positioned between the functional blocks and the shared memory to enhance data access flexibility and efficiency, supporting mixed-grained data prefetching and programmable dynamic data allocation. The lightweight 32-bit integer
RISC-V Snitch core \cite{zaruba2020snitch} orchestrates the functional blocks and data streamers through RISC-V Configuration and Status Register (CSR) instructions, while a DMA core manages off-chip data movement.
\vspace{-0.2cm}
\subsection{3D Spatial Data Reuse by GEMM Core}
\label{sec:arch_gemm}
As shown in Figure \ref{fig:chip_arch} (c), the core component of the GEMM core is a 3D spatial array consisting of 512 multiply-and–accumulate (MAC) units organized in an 8×8×8 manner \cite{yi2025opengemm}. Each basic processing element in this array is a dot-product unit (Dot-ProdU) that computes the dot product of two 8-element vectors. The Dot-ProdUs are further arranged in an 8×8 2D topology.
This architecture exploits 3D spatial data reuse: vectors from the input and weight matrices are broadcast horizontally and vertically across the Dot-ProdU array, while within each Dot-ProdU, eight partial products are combinationally accumulated to produce a single output. 
The GEMM core adopts an output-stationary dataflow to maximize temporal data reuse, avoiding frequent access of the high data precision partial sum and the output result. A built-in hardware loop controller generates the necessary control signals to clear the accumulation registers according to the RISC-V core Snitch programmed matrix dimensions.

% \subsection{Flexible Shared Memory Access}
% \label{sec:arch_data}

% \subsubsection{Flexible Data Access}
% To flexibly support diverse data access patterns required by various GEMM and Conv2D workloads, \gemmchipname integrates a 6-D AGU in the streamer A and a 3-D AGU in the streamer B. The 6-D AGU can generate the address streams for programmable 6-D affine access, effectively supporting \textit{implicit im2col} \cite{zhou2021characterizing} for all types of convolutions.

\subsection{Mixed-Grained Data Pre-fetching by Flexible Data Streamers}

\begin{figure}[t]
	\centering
	\includegraphics[width=1\linewidth]{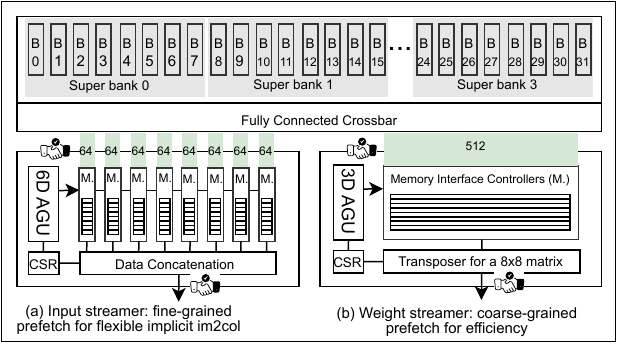}
	\caption{Flexible data streamer architecture and mixed-grained data prefetching mechanism: (a) fine-grained data prefetching and (b) coarse-grained data prefetching.}
	\label{fig:data_prefetch}
    \vspace{-0.4cm}
\end{figure}

As shown in Figure \ref{fig:chip_arch} (b), seven flexible data streamers \cite{yi2025datamaestro} serve as key components that deliver continuous operand data streams to the functional blocks within \gemmchipname, efficiently interfacing with the shared memory structure. As illustrated in Figure \ref{fig:data_prefetch}, each streamer primarily consists of a multi-dimensional Address Generation Unit (AGU), Memory Interface Controllers (MICs), and data FIFOs. 
The AGU’s loop dimensionality is determined by the complexity of the required data access pattern.
To flexibly support the diverse access patterns of various GEMM and Conv2D workloads, \gemmchipname employs a 6-D AGU in the input streamer and a 3-D AGU in the weight streamer. The 6-D AGU generates the address streams for programmable 6-D affine access, effectively supporting the strided data access pattern of \textit{implicit im2col} \cite{zhou2021characterizing} for all convolution types, covering arbitrary stride, kernel size, input channel, output channel, and the block-wise data access pattern of GEMM operations with varying matrix sizes. The base address pointers, loop bounds, and strides \cite{yi2025datamaestro} for the multi-dimensional affine address generation are programmed to the data streamers by the Snitch core through CSR registers.

% \xyi{[also mention narrow and wide somewhere]}
To mitigate bank contention within the shared memory \cite{genc2021gemmini, yi2025datamaestro}, a mixed-grained data prefetching (MGDP) mechanism is implemented by inserting data FIFOs into the streamers’ access channels, which are the smallest units responsible for memory interactions, as shown in Figure~\ref{fig:data_prefetch}. Both the input and weight streamers contain eight-depth FIFOs each. When a FIFO is not full, the MIC proactively prefetches the next data segment based on the generated address stream, effectively hiding memory access latency from the GEMM core’s perspective. 
Considering that Conv2D workloads require strided data accesses with finer granularity when fetching feature maps \cite{zhou2021characterizing}, the input streamer adopts a 64-bit channel width (Figure~\ref{fig:data_prefetch} (a)). In contrast, the weight streamer operates with coarser access granularity based on the weight-fetch patterns of Conv2D and GEMM workloads, as long as the weight operands are arranged in an appropriate data layout~\cite{zhou2021characterizing, yi2025datamaestro}. Thus, the weight streamer employs a 512-bit channel width to efficiently fetch the required data in parallel by accessing a super bank, which is a 512-bit-wide bank formed by combining eight shared-memory banks (Figure~\ref{fig:data_prefetch} (b)).
Thanks to the output-stationarity of the GEMM core, which maximizes the temporal reuse of output results within the array, \gemmchipname incorporates 1-depth data FIFOs in both the partial sum and output streamers.

% \subsubsection{On the Fly Data Manipulation}
\subsection{Programmable Dynamic Memory Allocation}
 % by Shared Memory Structure
 
\begin{figure}[t]
	\centering
	\includegraphics[width=1\linewidth]{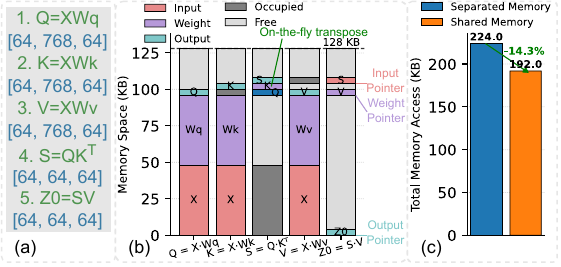}
	\caption{Dynamic memory allocation example for Multi-Head Attention (MHA). (a) Detailed MHA operation in BERT-Base model with one head and token size=64. (b) Memory allocation for each operand in the MHA computation sequence. (c) Saved memory access count.}
	\label{fig:shared_mem}
    % \vspace{-0.5cm}
\end{figure}

The shared memory structure and flexible data streamers jointly enable programmable dynamic memory allocation (PDMA), allowing the on-chip data memory to be flexibly (re)partitioned and (re)used according to workload characteristics and data access patterns. Figure~\ref{fig:shared_mem}~(a) illustrates the multi-head attention (MHA) computation sequence and matrix dimensions of the BERT-Base model, while Figure~\ref{fig:shared_mem}~(b) shows an example of dynamically partitioning \gemmchipname’s shared memory and mapping this MHA sequence onto it. By leveraging a unified memory space for input, weight, and output operands, along with dynamic base pointer updates in the corresponding data streamers, no data transfers between separate input/weight/output buffers and off-chip memory are required. This approach reduces total data access counts by 14.3\%, compared to a separated-memory architecture with fixed data dispatchers where the data transfers between separated memory and off-chip memory are required, as shown in Figure~\ref{fig:shared_mem}~(c).
In addition, a built-in data transposer, as shown in Figure \ref{fig:data_prefetch} (b), within the weight streamer performs the $K^{T}$ operation on the fly, further reducing redundant memory access compared to using a dedicated data transposer~\cite{genc2021gemmini}.

% \textbf{Figure 3: show the benefit originated in the shared memory architecture. MHA data allocation demo, use Bert-Base one head, to show all data fit in the shared memory, no L3 write back, and dynamic I/W/O allocation (previous output becomes current input).}

\subsection{Time-Multiplexing}
\label{sec:arch_time_mux}

\gemmchipname leverages time-multiplexing techniques to reduce module area cost with negligible performance degradation.
The quantization SIMD unit takes in the GEMM core’s 32-bit outputs and converts them to 8-bit precision.
By exploiting the GEMM core’s output-stationary dataflow, the SIMD unit instantiates only eight quantization PE lanes. These PEs perform quantization and activation on the GEMM core’s outputs, processing 64 results over eight cycles using a hardware loop unroller, as illustrated in Figure \ref{fig:chip_arch} (d). Evaluation results show that the time multiplexed SIMD unit incurs only $0.7$\% performance loss on the ResNet50 workload while achieving a $4.92\times$ reduction in SIMD area, compared with a 64-lane SIMD design. 
Besides, the partial-sum and output streamers are also time-multiplexed when interacting with the shared memory, effectively halving the access ports of the fully connected crossbar, as shown in the middle of Figure \ref{fig:chip_arch} (b). \gemmchipname prioritizes partial-sum reads over output writes, since output data are generated only after the partial sums are forwarded to the GEMM core. Evaluation results show that this optimization results in merely $0.02$\% performance loss on the ResNet50 workload while reducing the crossbar area by $1.46\times$.

% \subsubsection{Time-Multiplexed SIMD}
% \subsubsection{Time-Multiplexed Data Ports}

% \begin{figure}[t]
% 	\centering
% 	\includegraphics[width=1\linewidth]{img/gemm_chip_figure-f4_time-mux benifits.pdf}
% 	\caption{(a) Performance evaluation and area saving for SIMD time-multilex. (b) Performance evaluation and area saving for data streamer C and D time-multilex.}
% 	\label{fig:time_mux}
%     \vspace{-0.6cm}
% \end{figure}

% \textbf{Figure 4: benefit of time multiplex. (a) SIMD with time-based reuse, reducing the 64 PEs to 8 PEs. A figure to show the performance drop is minimal, and the area saved is a lot. (b) Data movers C and D are merged. A figure to show the performance drop is minimal, and the area saved is a lot.}

\subsection{Auxiliary Modules}
\label{sec:auxi}

\gemmchipname integrates a dedicated data reshuffler to support various DNN operations that require data layout transformations. The data layout transformation unit reorganizes data from its original layout into a new format that meets the requirements of the GEMM core or minimizes bank contention. For instance, it can transform a row-major layout into a blocked row-major format for GEMM operation's input matrix, or convert an HWC layout into a C/8HWC8 format for Conv2D's feature maps \cite{yi2025datamaestro}, enabling the input streamer to access data efficiently and correctly.
In addition, the maxpool unit features eight parallel comparison lanes and can be configured to perform max-pooling operations with arbitrary window sizes in a sequential manner.

% \subsubsection{MaxPool Unit}
% \subsubsection{Data Layout Transformation Unit}
% \subsubsection{Transpose Unit}

%% file: content/3-exp.tex
\section{Measurement Results}

\begin{figure}[t]
	\centering
	\includegraphics[width=0.95\linewidth]{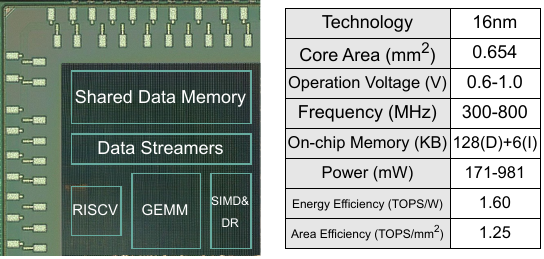}
	\caption{Chip micrograph and specification.}
	\label{fig:die_photo}
    % \vspace{-0.5cm}
\end{figure}

% \subsection{Chip Summary}
Figure \ref{fig:die_photo} shows the die photo and chip specification of \gemmchipname. The chip is fabricated in 16nm FinFET technology with a total area of $0.654$mm\textsuperscript{2}. It operates at supply voltages ranging from $0.6$-$1$V, with a frequency of $300$-$800$MHz.

\subsection{Performance Evaluation}
\label{sec:perf}

\begin{figure}[t]
	\centering
	\includegraphics[width=1\linewidth]{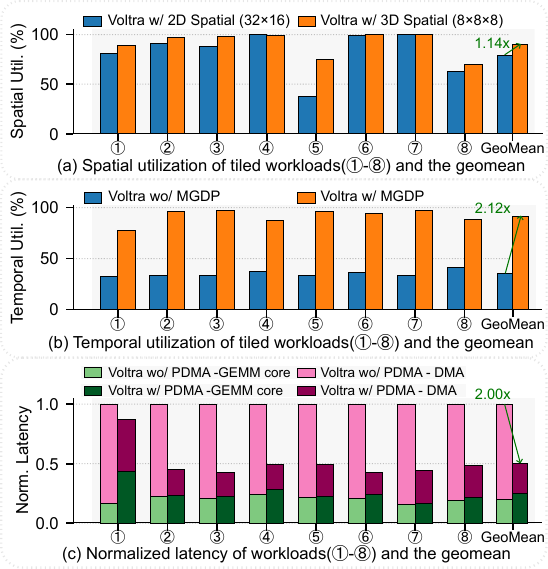}
	\caption{Performance evaluation result across diverse workloads: MobileNetV2 (\ding{172}), ResNet50 (\ding{173}), ViT-B (\ding{174}), PointNeXt (\ding{175}), LSTM (\ding{176}), BERT-Base (\ding{177}, input token size = 512), LLaMA3.2-3B prefill stage (\ding{178}, input token size = 256) and LLaMA3.2-3B decode stage~(\ding{179}, input token size = 256) and their geomean. (a) The spatial utilization benefits of the 3D spatial array compared with a 2D spatial array. (b) The temporal utilization benefits of the mixed-grained data prefetching (MGDP) technique. (c) The total latency benefits of the programmable dynamic memory allocation (PDMA) mechanism. }
	\label{fig:sys_perf}
    % \vspace{-0.5cm}
\end{figure}

We conduct an extensive performance evaluation of \gemmchipname across a diverse set of CNN, RNN, and transformer workloads, with the results presented in Figure~\ref{fig:sys_perf}. In each subfigure, the right bars represent the performance of \gemmchipname, while the left bars correspond to the baseline results with each proposed feature individually disabled. Since the $128$KB on-chip memory of \gemmchipname is insufficient to accommodate the full operands of these workloads, we apply layer-wise tiling, where each layer is partitioned to fully exploit the GEMM core’s output-stationary dataflow \cite{mei2021zigzag}. The utilization metrics in Figure~\ref{fig:sys_perf} are measured within these tiled layer blocks, while the total latency metric additionally accounts for off-chip data movement\footnote{Off-chip data movement cycles are simulated using a cycle-accurate RTL model.} of tiles during the whole workload execution.

As shown in Figure~\ref{fig:sys_perf}~(a), \gemmchipname achieves $69.71$\%-$100$\% spatial utilization across these eight workloads, representing up to $2.0\times$ improvement over a 2D spatial array design. The LLM decode stage workload exhibits a lower spatial utilization of $69.71$\%  due to the mismatch between the workload size (where a lot of GEMV operations occur) and the 3D spatial array shape of \gemmchipname.
Benefiting from the MGDP technique, \gemmchipname attains as high as $76.99$\%-$97.32$\% temporal utilization across these workloads, delivering a $2.12$-$2.94\times$ improvement over a plain shared memory architecture, which suffers from severe bank contention, as illustrated in Figure~\ref{fig:sys_perf}~(b). Furthermore, the PDMA mechanism enables larger tile sizes, effectively reducing off-chip communication, compared to a separate memory architecture. 
% Consequently, as shown in Figure~\ref{fig:sys_perf}~(c), even though the GEMM core computation cycles increase slightly due to the higher temporal utilization in the separated memory, the DMA data movement cycles in shared memory are reduced, leading to an overall $1.15$-$2.36\times$ improvement in total latency.
Consequently, as shown in Figure~\ref{fig:sys_perf}~(c), although the GEMM core computation cycles increase slightly due to the higher temporal utilization in the separated-memory configuration, the adoption of the PDMA mechanism with a shared memory significantly reduces the DMA data movement cycles, resulting in an overall $1.15$–$2.36\times$ reduction in total latency.
% \subsubsection{Spatial Utilization}
% \subsubsection{Temporal Utilization}
% \subsubsection{Latency}

\subsection{Efficiency Evaluation}

\begin{figure}[t]
	\centering
	\includegraphics[width=1\linewidth]{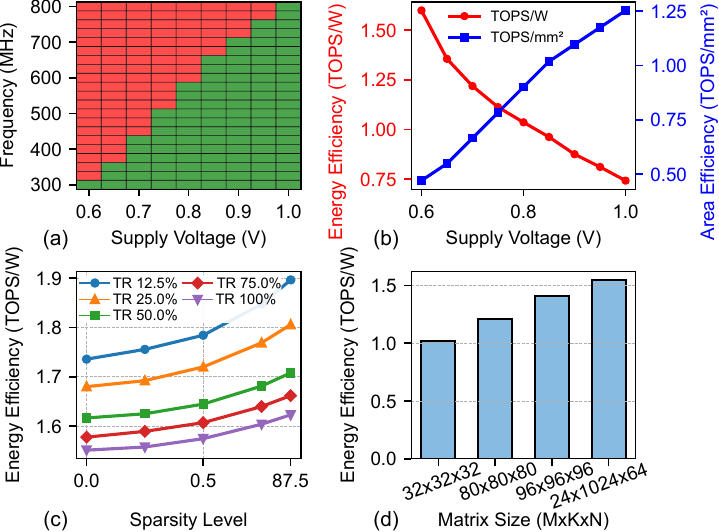}
	\caption{The chip measurement results of (a) shmoo plot, (b) the energy efficiency and area efficiency at different supply voltages, (c) energy efficiency at different weight sparsity and input toggle rates (TR), (d) the effective energy efficiency for different matrix sizes.}
	\label{fig:shmoo}
    % \vspace{-0.4cm}
\end{figure}

Figure~\ref{fig:shmoo}~(a) presents the shamoo plot of \gemmchipname, while Figure~\ref{fig:shmoo}~(b) illustrates the trends of system energy efficiency and area efficiency as the supply voltage increases. The results in Figure~\ref{fig:shmoo}~(b) are obtained using a fully dense GEMM workload with $M=N=K=96$. A peak system energy-efficient point of $1.60$ TOPS/W is achieved at $0.6$V and $300$MHz. The peak system area efficiency reaches $1.25$~TOP/mm\textsuperscript{2} at $1.0$V and $800$MHz. Figure~\ref{fig:shmoo}~(c) depicts the sparsity characteristics of \gemmchipname, showing the energy efficiency under varying weight sparsity levels and input toggle rates.
Figure~\ref{fig:shmoo}~(d) illustrates the energy efficiency trend for GEMM workloads with different matrix sizes. Larger matrices enable greater data reuse, with the $K$ dimension in particular enhancing temporal data locality due to the GEMM core's output-stationary dataflow, thereby improving the overall system efficiency.

% , GOPSvsTOPS/W, GOPSvsTOPS/mm2.

% Area and Power breakdown (post synthesis) \textbf{Figure 7}

\subsection{State-of-the-Art Comparison}
\label{sec:sota}

% diana, RBE, , this work

Table~\ref{tab:sota_compare} summarizes the comparison between \gemmchipname and several state-of-the-art (SotA) hardware accelerator chips, including DIANA~\cite{houshmand2022diana}, REB~\cite{conti2024marsellus}, Ayaka~\cite{qin2024ayaka}, and Cygnus~\cite{jain2025cygnus}.
\gemmchipname is the only design that efficiently supports a wide range of workloads, from CNNs to Transformers. As discussed in Section \ref{sec:perf}, it maintains high utilization across diverse workloads, enabled by the 3D spatial data reuse GEMM core and the efficient, flexible shared memory access architecture, where the mixed-grained data prefetch and programmable dynamic memory allocation are realized.
When evaluated on a fully dense GEMM workload, \gemmchipname achieves competitive system energy efficiency.
Moreover, it exhibits the smallest chip area and the highest system area efficiency among the compared designs, largely attributed to the time-multiplexing techniques employed in the SIMD unit and the partial sum and output data streamers.

\begin{table}[htbp]
	\begin{threeparttable}
		\centering
		\scriptsize
		\caption{Chip measurement result summary and comparison of SotA DNN accelerators.}
		\label{tab:sota_compare}
		\setlength{\tabcolsep}{2pt}
		\begin{tabular}{c}
        \includegraphics[width=1\linewidth]{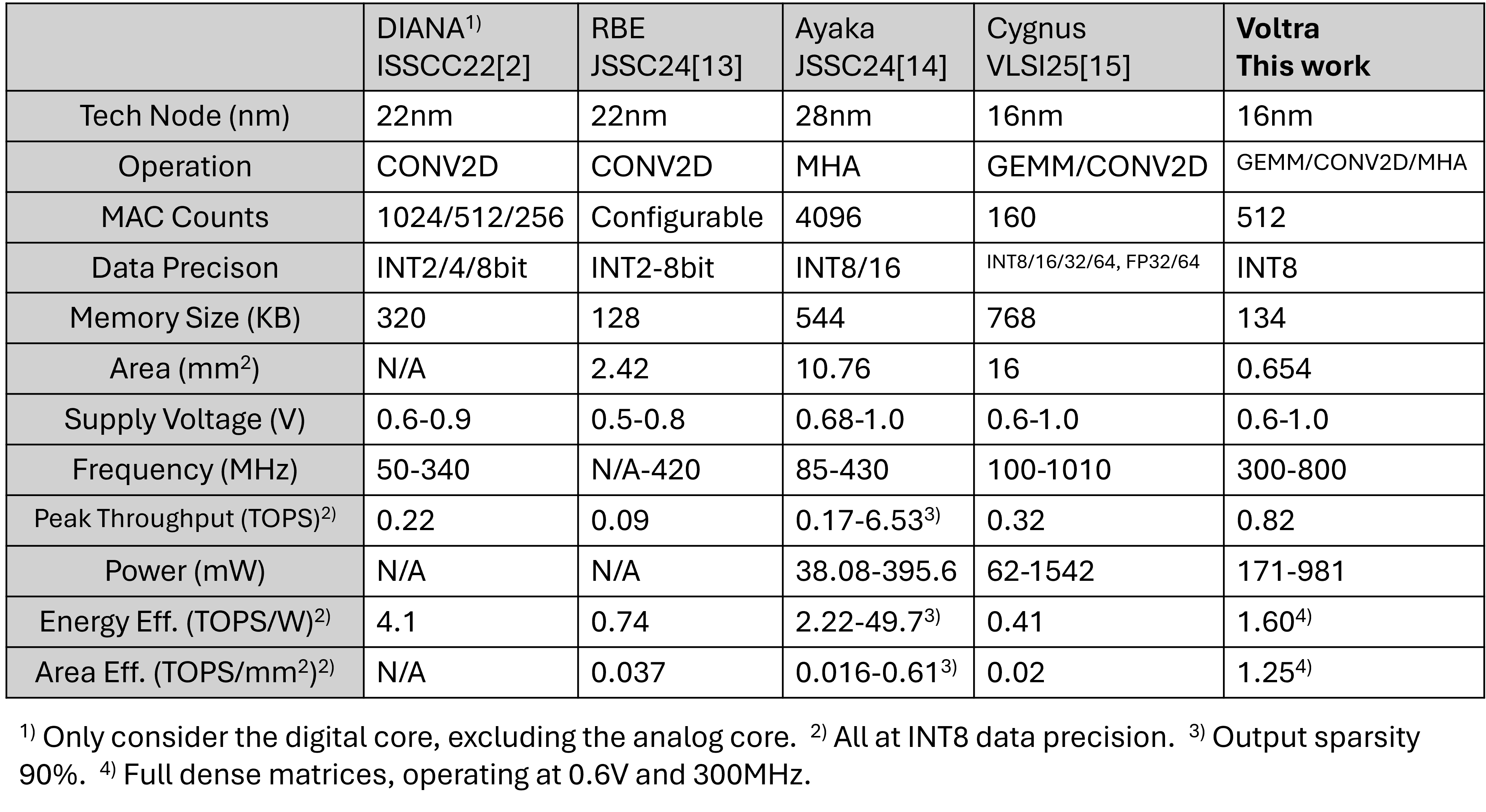} \\
		\end{tabular}
		% \begin{tablenotes}
		% 	\scriptsize
		% 	\item[1] Measured at specified Vdd/freq.
		% \end{tablenotes}
	\end{threeparttable}
    % \vspace{-0.4cm}
\end{table}

%% file: content/4-conclusion.tex
\section{Conclusions}
\label{sec:conclusions}

% In this work, we present a high-utilization DNN accelerator that employs a novel 3D spatial data reuse strategy along with a flexible and efficient shared memory access architecture. The 3D spatial dataflow enhances spatial utilization by enabling efficient spatial data reuse across three dimensions, while the flexible shared memory access architecture incorporates mixed-grained hardware data pre-fetching to boost temporal utilization and programmable dynamic data allocation to reduce off-chip data movement cost. 
% Fabricated in 16nm technology, our chip achieves a peak system energy efficiency of $1.60$ TOPS/W and a system area efficiency of $1.25$ TOPS/mm\textsuperscript{2} when running a full dense GEMM workload. The measurement results demonstrate that our design is competitive with state-of-the-art solutions while maintaining high utilization across diverse workloads.

In this work, we present a high-utilization DNN chip that employs 1) a novel 3D spatial data reuse strategy to enhance spatial utilization, along with 2) mixed-grained hardware data pre-fetching to boost temporal utilization and 3) programmable dynamic data allocation to reduce off-chip data movement cost in a flexible and efficient shared memory access architecture. 
Fabricated in 16nm technology, our chip achieves a peak system energy efficiency of $1.60$ TOPS/W and a system area efficiency of $1.25$ TOPS/mm\textsuperscript{2} when running a full dense GEMM workload. The measurement results demonstrate that our design is competitive with state-of-the-art solutions while maintaining high utilization across diverse workloads.

%% file: content/5-ack.tex
\section{Acknowledgments}
\label{sec:acknowledgments}

This project has been partly funded by the European Research Council (ERC) under grant agreement No. 101088865, the European Union's Horizon 2020 program (CONVOLVE) under grant agreement No. 101070374, the Flanders AI Research Program, and KU Leuven.